# Effect of spark plasma sintering on the superconducting properties of Sm-based oxypnictide

Mohammad Azam[1], Tatiana Zajarniuk[2], Konrad Kwatek[3], Paolo Mele[4], Shiv J. Singh[1†]

[1]*Institute of High Pressure Physics (IHPP), Polish Academy of Sciences, Sokołowska 29/37, 01-142 Warsaw, Poland*

[2]*Institute of Physics, Polish Academy of Sciences, Aleja Lotników 32/46, 02-668 Warsaw, Poland*

[3]*Faculty of Physics, Warsaw University of Technology, Koszykowa 75, 00-662 Warsaw, Poland*

[4]*College of Engineering, Shibaura Institute of Technology, 307 Fukasaku, Minuma-ku, Saitama 337-8570, Japan*

[†]Corresponding author: Shiv J. Singh

Email: sjs@unipress.waw.pl

https://orcid.org/0000-0001-5769-1787

# Abstract

We optimize the superconducting properties of Sm-based oxypnictide (Sm1111: $SmFeAsO_{0.80}F_{0.20}$) by using the Spark Plasma Sintering (SPS) technique under various synthesis conditions, including heating temperatures ranging from 600 to 1000 °C for durations of 5 to 30 minutes at the applied pressure of 45 MPa. All prepared bulks are characterized by structural and microstructural analysis as well as transport and magnetic measurements to conclude our findings. $SmFeAsO_{0.80}F_{0.20}$ bulks are also prepared using the conventional synthesis process at ambient pressure (CSP) and the high gas pressure and high temperature (HP-HTS) methods at 500 MPa, which exhibit a superconducting transition temperature ($T_c$) of 53–54 K. Interestingly, the SPS process of $SmFeAsO_{0.80}F_{0.20}$ increases the sample densities up to 97-98% and confirms the optimized synthesis conditions of 900°C for 5-10 min; however, the increased sintering temperature or duration reduces $T_c$ due to the possible evaporation of lighter elements, particularly fluorine. Furthermore, the SPS technique is unable to reduce the observed impurity phases for the Sm1111, which is similar to the CSP and HP-HTS processes. A slight increment in the $J_c$ by the SPS process is observed due to the enhancement of sample density. A comparative analysis of Sm1111 superconductors prepared by SPS is performed with CSP and HP-HTS processes, suggesting that an increased sample density is ineffective on the superconducting properties in the presence of the impurity phases. This finding can be beneficial for the fundamental and applied research of iron-based superconductor (FBS).

**Keywords:** Iron-based superconductors, Spark plasma sintering, Critical current density, Impurity phases, Critical transition temperature

# Introduction:

The discovery of iron-based superconductors (FBS) in 2008 provides a second high-$T_c$ material [1]. The chemical diversity of these materials has revealed the existence of over 100 compounds, each with potential for electron and hole doping. The phase diagram that results from this class of superconductors is both intriguing and exciting from a scientific standpoint, and as such, further research is required to facilitate a comprehensive understanding [2-5]. FBS was discovered with F-doped LaOFeAs, which provides the superconducting transition ($T_c$) of 26 K and belongs to the 1111 ($RE$FeAsO, where $RE$ = Rare Earth) family) [1]. Following this pioneering work, the superconducting transition increased up to 58 K in F-doped SmFeAsO (Sm1111) [6,7] which is the highest $T_c$ value reported for FBS. The 1111 family can be a strong contender of cuprate superconductors for the applications due to high $T_c$, high upper critical field ($H_{c2}$~100 T), and high critical current density ($10^6$ A/cm$^2$ at 5 K) [4]; however, the growth of these materials is one of the basic challenges, and as a result, this family is not explored much [5]. Due to the easy synthesis process, most research works are devoted to the 122 ($A$Fe$_2$As$_2$, $A$ = Ba, K) and 11 (FeSe) families [5]. Since the parent compound of the 1111 family provides the structural and magnetic transition around 150 K [7], but by a suitable doping, the superconductivity can be observed [2,5]. Various kinds of doping, such as fluorine (F) [6], cobalt (Co) [8], nickel (Ni) [9], phosphorus (P) [10] and iridium (Ir) [11], thorium (Th) [12] have been reported for the 1111 family to induce and enhance their superconducting properties. Among these doping, fluorine substitution at oxygen sites is most effective and provides the highest transition temperature of 58 K [6,7, 13]. Since fluorine is a lighter element, it is also difficult to reduce its evaporation during the high-temperature synthesis process. Hence, one of the significant challenges is the synthesis of the high-quality 1111 samples with high sample density, which directly affects their superconducting properties, especially the critical transition temperature ($T_c$) and current density ($J_c$). Over the past 16 years since the discovery of FBS, most of the studies have employed the conventional synthesis process at the ambient pressure (CSP) [5] and suggest that CSP is inadequate to address the fundamental sample issues of FBS [2, 5,14,15] necessitating the implementation of novel methodologies such as high-pressure growth methods. However, very few investigations [16,17] had been performed using high-pressure growth techniques for the 1111 family. Recently, the high-gas pressure and high temperature synthesis (HP-HTS) [18] method was applied to grow F-doped Sm1111 bulks under different pressures and durations to optimize the synthesis parameters, suggesting that 500 MPa for one hour is the best condition at a synthesis temperature of 900 °C. The optimum F-doped Sm1111 bulks exhibited a slightly enhanced critical current density ($3 \times 10^3$ A/cm$^2$) compared to that (~$10^3$ A/cm$^2$) of CSP-processed samples, while exhibiting a nearly same transition temperature ($T_c$~53-54 K) and a slightly improved sample density (~58%) [19]. It is noteworthy that these Sm1111 manifest the same type and quantity of impurity phases (SmOF and SmAs), regardless of the preparation method employed, whether HP-HTS or CSP. Hence, the HP-HTS method cannot reduce the observed impurity phases during the phase formation of Sm1111, but it results a slight improvement in the superconducting properties of Sm1111 [19] in comparison to CSP. Karpinski et al. [20,21] reported the high-pressure growth of various 1111 single crystals; however, the crystal size was very tiny (100-300 μm), rendering them unsuitable for various kinds of characterizations. These studies suggest that high-pressure growth for FBS is effective; however, further research employing various pressure techniques is urgently required.

Spark Plasma Sintering (SPS) is a pressure-assisted sintering method that employs pulsed electric currents to facilitate rapid densification of materials by minimizing porosity [22-25]. This technique has been demonstrated to offer several advantages, including reduced processing times, the capacity to operate across a range of sintering temperatures, and enhanced sample density, which frequently attains levels of up to 99-100%. Apart from high-pressure growth processes such as the HP-HTS method, SPS can be a promising method for addressing the basic challenge of FBS [26]. This technique is applied to K-doped BaFe$_2$As$_2$(Ba122), resulting in an enhancement of the sample density exceeding 90% and an increase in the critical current density $J_c$ by one order of magnitude at 5 K [27]. Similarly, CaKF$_4$As$_4$ (1144) bulks were also prepared using the SPS method, achieving a sample density of 96.2%, and enhancing $J_c$ by one order of magnitude; however, eliminating the impurity phases remained unachieved [28]. Comparable findings by SPS were also observed for other FBS families, such as Graphene-doped FeSe$_{0.5}$Te$_{0.5}$ depicted a high critical field ($H_{c2}$) with a density of 90% [29]; NaFeAsO$_{0.75}$F$_{0.25}$ with a higher density of 96%, showed a slightly lower onset $T_c$ [30]; Co-doped BaFe$_2$As$_2$ improved the sample density up to 80% density [26]; and recently, Ishida et al. reported the improved $J_c$ of 127 kA cm$^{-2}$ at 4.2 K via spark plasma texturing techniques for the 1144 family [31]. These studies motivated us to apply the SPS method to F-doped Sm1111, which provides the highest $T_c$ of 58 K for FBS. Furthermore, there are currently no reports on the growth optimization of the Sm1111 family using the SPS approach and its impact on their superconducting properties.

In this paper, we have optimized the synthesis process of F-doped Sm1111 bulks using the SPS method by preparing several bulks under varying growth parameters, including sintering temperature and duration. A comprehensive characterization has been conducted for all prepared bulks to reach the final conclusion. F-doped Sm1111 materials are also prepared under the optimal conditions using the CSP and HP-HTS methods for a comparative analysis with Sm1111 samples produced by SPS. Sm1111 bulks processed by SPS have enhanced the sample density, which is almost twice that of the CSP and HP-HTS, but the observed superconducting properties are almost the same as those of HP-HTS. A comprehensive analysis of CSP, HP-HTS, and SPS processes for F-doped Sm1111 bulks suggests that the presence of impurity neutralizes the impact of increased sample density on the superconducting properties of a material. This work represents the first detailed, comprehensive investigation of Sm1111 superconductors by three different growth processes, which can be fruitful for the exploration of FBS materials. Since the optimal fluorine doping content for SmFeAsO is 20% [6], we have selected this fluorine doping content (20%) for optimizing the synthesis process of Sm-based oxypnictides through SPS and high-pressure growth processes. This composition exhibits the highest superconducting transition temperature ($T_c$) with the minimal impurity phases [6]. It is well established that doping levels exceeding 20% fluorine for Sm-based oxypnictides result in a nearly constant superconducting transition temperature ($T_c$), while simultaneously causing a rapid increase in the impurity phases. Therefore, the $SmFeAsO_{0.80}F_{0.20}$ composition is well-suited for investigating the fundamental physics and is a potential candidate for future applications.

## Experimental details

Polycrystalline $SmFeAsO_{0.80}F_{0.20}$ was prepared using initial precursors, including Sm powder (99.9 %), As chunks (99.99%), Fe powder (99.99%), $Fe_2O_3$ powder (99.85%), and $FeF_2$ powder (99%), as discussed elsewhere [19,32,33]. First, SmAs were initially prepared by reacting samarium and arsenic powders at 500 °C for 15 hours as a precursor powder. These initial precursors were used to synthesize bulk $SmFeAsO_{0.80}F_{0.20}$ following a one-step solid-state reaction method. According to the stoichiometric formula $SmFeAsO_{0.80}F_{0.20}$, the starting materials were mixed and ground using a mortar and pestle, and thereafter compressed with a hydraulic press at about 200 bars to form disk-shaped pellets with a diameter of 10 mm. The pellets were subsequently placed in a tantalum (Ta) tube, serving as a crucible, and sealed within an evacuated quartz tube, which was then heated to 900 °C for 45 hours [19]. All growth processes were performed in an inert gas glove box with oxygen and moisture levels below < 1 ppm, in the batches of approximately 6 grams. The bulk material synthesized by this process is referred to as the parent compound (P). This prepared sample was cut into a rectangular-shaped piece for the physical property measurements, while the remaining samples were used for the subsequent SPS process. The parent P pellets were ground to an average size of 100-200 microns to achieve particle homogenization and improve the densification of a pellet during the SPS process. The approximately 0.5 grams of reground $SmFeAsO_{0.80}F_{0.20}$ powder were used for each SPS experiment. The powders were contained within a graphite die featuring an internal diameter of 10 mm, sealed by two graphite punches. The pellet underwent sintering in the die utilizing an SPS LABOX-210 (NJS Co., Ltd., Yokohama, Japan) Sinterland, maintained at a constant pressure of 45 MPa in a vacuum environment [34]. The sample was heated at three distinct temperatures for varying durations: (a) 600 °C for 5-10 min, (b) 900 °C for 5-30 min, and (c) 1000 °C for 5-10 min for various Sm1111 experiments, as mentioned in Table 1. Following the SPS process, the sample was allowed to cool down to room temperature prior to the release of the applied pressure. The pellet was extracted from the graphite dies and subsequently polished to eliminate any remnants of the protective graphite foil. The resulting pellet measured around 1 mm in thickness and 10 mm in diameter. Furthermore, we also prepared HIP bulk using the optimized synthesis conditions of 900 °C for 1 hour at 500 MPa using the HP-HTS method [18], as previously reported elsewhere [19]. The $SmFeAsO_{0.80}F_{0.20}$ bulks synthesized via CSP and HP-HTS techniques are referred to as Parent P and HIP, respectively. Table 1 provides comprehensive details regarding all SPS bulks, including their synthesis conditions and sample codes, which were rigorously evaluated using various techniques to examine their influence on the material's superconducting properties. We have calculated the density of a sample using its mass and the dimensions of its disk-shaped bulk, employing the formula: density = mass/volume. The calculated density for each sample is compared with the reported theoretical density of the SmFeAsO (Sm1111) phase, which is 7.1 g/cm³ [35], to determine the relative density, as similar to previous reports [28, 33]. Table 2 provides the synthesis parameters for the best SmFeAsO$_{0.80}$F$_{0.20}$ bulks processed by SPS and HP-HTS method with the parent sample.

The X-ray diffraction (XRD) method was employed to assess the phase purity of powder samples. Data collection was carried out using a Philips X'Pert Pro diffractometer within the 2θ range of 10° to 90°, with a step size of 0.013° and a counting time of 300 seconds per step, utilizing CuKα radiation (λ = 1.5418 Å) at 30 mA and 40 kV. Profile and phase purity analysis and lattice parameter calculations were conducted using the standard diffraction database PDF4+ 2024, provided by the ICDD database. A Zeiss Ultra Plus field-emission scanning electron microscope (FESEM) with Bruker Quantax 400 EDS microanalysis equipment was employed for comprehensive microstructural analysis and elemental mapping. Vibration-sample magnetometry (VSM) integrated with a Physical Property Measurement System (PPMS) is employed to assess magnetic characteristics within a temperature range of 5-60 K under an applied magnetic field of up to 9 T. Magnetic susceptibility measurements were conducted in both field-cooled (FC) and zero-field-cooled (ZFC) modes under an applied magnetic field of 20 Oe on the rectangular-shaped samples. The temperature dependence of the resistivity was measured using the four-probe method at various DC currents of 5 mA, 10 mA, and 20 mA using a closed-cycle refrigerator (CCR) within the temperature range of 7-300 K in the absence of a magnetic field. The rectangular-shaped bulks are used for the transport and magnetic measurement of each sample having different sample dimensions.

## Result and discussion

### 1. Structural analysis

The purity and crystallinity of the synthesized SmFeAsO$_{0.80}$F$_{0.20}$ polycrystalline samples were examined using powder X-ray diffraction (XRD) measurements. Figure 1 presents the room temperature XRD patterns of all Sm1111 bulks prepared by the SPS method under various synthesis conditions (refer to Table 1), alongside the parent P and HIP bulks. The crystal structure of all samples analyzed in the XRD patterns corresponds to the tetragonal ZrCuSiAs-type structure. The observed phases are indexed according to this structure, defined by the space group P4/nmm as reported for the Sm1111 [1,6,33] compound. The parent P contains the impurity phases of SmOF (Sm$_2$O$_3$) (6-7%) and SmAs (2%), which is consistent with previous reports [6,19]. The HIP bulks produced under optimized growth conditions exhibit comparable quantities and types of impurity phases with the parent P sample. This indicates that a synthesis pressure of 500 MPa was unable to reduce the impurity phases for F-doped Sm1111, as observed for the HIP sample [19]. Interestingly, all SPS bulks exhibit almost the same quantity and type of impurity phases as those identified in the parent P and HIP samples, as detailed in Table 3. Figure 1(b) presents an enlarged view of the main peak (102) of the tetragonal phase for all these samples. The samples SPS-1 and SPS-2 have the peak (102) position comparable to that of the parent and HIP samples, suggesting almost the same fluorine content within the Sm1111 lattice. For SPS-3, SPS-4, and the samples for longer sintering durations of 20 and 30 minutes (SPS-5 and SPS-6), there has been a slight shift of this peak position to the low $2\theta$ angle. The shift of the peak (102) is more pronounced in the SPS-7 and SPS-8 samples that were heated to 1000 °C for 5 and 10 minutes, respectively. Generally, fluorine (F) substitution at oxygen sites reduces the lattice parameters of SmFeAsO, as $F^-$ has a smaller ionic radius of approximately 1.33 Å compared to $O^{2-}$ at around 1.40 Å [36]. At optimal fluorine doping ($x$ = 0.2), the incorporation of $F^-$ results in lattice contraction compared to the undoped SmOFeAs, which typically causes diffraction peaks (such as 102) to shift toward higher $2\theta$ angle relative to the undoped sample due to decreased d-spacing. However, under non-optimal synthesis conditions such as higher temperatures or longer dwell times, as used for SPS-5 to SPS-8, a partial volatilization of fluorine can occur. This reduction in the fluorine content results in lattice expansion, causing the peak 102 to shift toward lower $2\theta$ angle (i.e., the left side compared to our parent sample), as noted for these SPS-5 to SPS-8 samples. A slight shift of the 102 peak toward the left side suggests a little amount of fluorine evaporation from the superconducting $SmFeAsO_{0.80}F_{0.20}$ lattice, as observed for the samples SPS-5 to SPS-8 (Figure 1(b)). This observation aligns with previously reported F-doped SmFeAsO, which exhibits an underdoped region of fluorine doping, specifically less than 20% [6] prepared by the CSP method and also as reported for $NdFeAsO_{0.75}F_{0.25}$ [30] bulks, which occur at elevated sintering temperatures exceeding 900 °C or with extended sintering durations surpassing 10 minutes during the SPS process.

Figures 1(c)-(e) illustrate the variation of the lattice parameters (*a* and *c*) and the lattice volume (*V*) across different SPS samples with parent P and HIP. The lattice parameters for the parent P sample are *a* = 3.928(7) Å and *c* = 8.497(9) Å, consistent with previous reports [6], [19], [33]. A slight variation in the lattice parameter '*a*' is observed for these samples (Figure

1(c)). In the case of the lattice '$c$,' the values for the parent P, HIP, SPS-1, SPS-2, SPS-3, and SPS-4 are nearly identical, whereas SPS-5 to SPS-8 exhibit a slight reduction compared to the parent P (Figure 1(d)). The observed variation in the lattice parameters '$c$' is minimal, which may be associated to slight changes in the actual fluorine content within the superconducting 1111 lattice. Figure 1(e) demonstrates the variation of the obtained lattice volume '$V$' for various Sm1111 samples prepared under different pressures and temperatures. A minor variation is noted among various SPS samples in comparison to the parent P; nonetheless, these variations are minimal and could be attributed to minor discrepancies in fluorine content within the different SPS bulks. While it is feasible, especially with prolonged sintering times and high temperatures during the SPS processes, the slight changes in lattice parameters complicates determining whether the fluorine levels have increased or decreased. Based on the peak shift observed in Figure 1(b), particularly for SPS-5, SPS-6, SPS-7, and SPS-8, it can be concluded that there is a slight variation in fluorine content in comparison to the parent F-doped Sm1111 [19], [30]. Since the selected fluorine content is optimal ($x = 0.20$), a small alteration in the lattice parameters remains undetectable due to a slight variation (~1-2%) in the actual fluorine contents as reported previously [37]. These analyses suggest that the SPS process performed at 900 °C for a duration of 5 to 10 minutes seems adequate for the synthesis of F-doped Sm1111 bulks. Consequently, prolonged sintering durations or higher heating temperatures (>900°C) are inappropriate for the SPS process.

## 2. Elemental mapping and analysis

Figure 2 presents the elemental mapping of the selected samples, while additional samples are depicted in the supplementary Figure S1 to analyze the distribution of constituent elements. Elemental mapping of these samples was conducted using energy-dispersive X-ray spectroscopy (EDAX). All bulks were meticulously polished with various grades of sandpaper inside a glove box, without the use of any lubricants. The samples generally exhibit degradation upon exposure to oil or alcohol; consequently, only sandpapers are utilized for the purpose of surface polishing. The parent and HIP samples exhibit the presence of Sm, As, O, and F, which are distributed nearly homogeneously within the samples, as depicted in Figures 2(i) and (ii). Specific locations demonstrate increased levels of Sm and O/F, suggesting the possible existence of the SmOF/$Sm_2O_3$ phase, as shown in Figures 2 and S1. Additionally, the presence of areas exhibiting high concentrations of Sm and As supports the presence of the SmAs phase, as evidenced by XRD analysis. In the HIP sample, a small amount of SmAs and SmOF ($Sm_2O_3$) is also detected in specific locations, was observed in the XRD analysis, similar to the findings of the earlier report [19]. The elemental mapping of the SPS-1 samples, prepared at 600 °C for 5 minutes, appears more porous and inhomogeneous compared to the parent P, while still maintaining the same types of impurity phases, as depicted in Figure S1(i). An increase in the sintering time, specifically 10 minutes at 600 °C (SPS-2), has resulted to a slight improvement in the uniform distribution of the constituent elements compared to SPS-1, as shown in Figure S1(ii). The samples SPS-3, SPS-4, and SPS-5 exhibit nearly identical distributions of constituent elements and confirm the presence of minor impurity phases, as illustrated in Figures S1(iii), 2(iii), and S1(iv), similar to the parent P and HIP. The extended sintering duration of 30 minutes appears to increase the inhomogeneity of the constituent elements in some areas of SPS-6 samples, as shown in Figure S1(v). The samples prepared at 1000°C for 5 and 10 minutes, namely SPS-7 and SPS-8, exhibit a nearly identical homogeneous distribution of the constituent elements, accompanied by the presence of impurity phases. The elemental mapping of all these samples reveals the presence of SmAs and SmOF/$Sm_2O_3$ phase, aligning with the previously discussed XRD data and analysis. This suggests that the SPS process at 900 °C for the extended sintering durations and at 1000 °C with different sintering times exhibits a slightly greater degree of inhomogeneity compared to SPS-3 and SPS-4. Nevertheless, the reduced sintering temperature of 600 °C is inadequate for effectively compacting the Sm1111 sample and achieving a homogeneous distribution of the constituent elements in comparison to other SPS samples. The elemental mapping suggests that a sintering temperature of 900 °C, along with a sintering duration of 5-10 minutes, is sufficient for achieving elemental homogeneity among the constituent elements.

## 3. Microstructural analysis

Microstructural analysis was performed on the polished surfaces of all Sm1111 bulks to investigate the intergrain connections and carry out the compositional analysis. Figure 3 illustrates the images for the selected samples: parent, HIP, SPS-4, and SPS-7 to understand the effects of CSP, HIP, and SPS processes on F-doped Sm1111 bulks. The remaining SPS bulks are presented in supplementary Figure S2(a)-(r). We acquired numerous images in backscattered electron (BSE) mode at different magnifications. The theoretical density of Sm1111 is reported to be 7.1 g/cm³ [35]; consequently, we have calculated the relative sample density using this value, as presented in Table 3. The images exhibit three distinct contrasts: bright, light grey, and black corresponding to $Sm_2O_3$/SmOF, $SmFeAsO_{0.80}F_{0.20}$, and pores, respectively. The black contrast may represent SmAs phase in specific cases. The parent material contains numerous small grains with nanopores between them, which reduces the grain connection and indicate a low sample density, as shown in Figure 3(a)-(c). The calculated density of 50% for the parent sample is consistent with this finding as mentioned in Table 3. The impurity phases, $Sm_2O_3$/SmOF and SmAs, are distributed uniformly throughout the sample area (Figure 3(a)) which is consistent with the above discussed elemental mapping. The HIP sample seems to be more compact than the parent P, which accounts for the higher sample density (~58%). In these samples, the impurity phase is identified as $Sm_2O_3$ alongside SmAs phase. Notably, the high-pressure processed Sm1111, referred to as HIP, exhibits an accumulation of the impurity phases rather than a homogeneous distribution within the sample, as illustrated in Figure 3(d-f), whereas other regions of the sample exhibit a clean superconducting Sm1111 phase characterized by numerous well-connected grain boundaries. The SPS-1 and SPS-2 samples prepared at 600 °C exhibit porosity and contain impurity phases of $Sm_2O_3$ and SmAs. Additionally, cracks are observed in certain areas, and the calculated density is approximately 42-43%, which is lower than that of the parent P. This analysis corresponds to the previously discussed elemental mapping. Nonetheless, SPS bulks processed at 900 °C for the duration of 5- and 10-min exhibit the localized impurity regions. At the same time, the majority of the samples maintain a clean superconducting phase, indicating effective sintering with minimal porosity and fine-grained materials, as demonstrated in Figure S2(g)-(i) for SPS-3 and Figure 3(g)-(i) for SPS-4. These samples exhibit better compactness and a higher degree of connectivity in their grain boundaries compared to SPS-1, SPS-2, the parent P and HIP samples. The observations corroborate the calculated density of 97-98% for these SPS-3 and SPS-4 samples (Table 3). Extended sintering durations, specifically 20 minutes (SPS-5) and 30 minutes at 900 °C (SPS-6), have accumulated impurity phases in certain regions. Concurrently, numerous small areas of the $Sm_2O_3$ phase are also evident, as illustrated in Figures S2(j)-(l). The extended sintering duration in the SPS process seems to have uniformly distributed the impurity phases throughout the sample, similar to the findings observed in the parent P sample. This suggests that a sintering duration of 20-30 minutes at 900 °C is suboptimal for the SPS process. Furthermore, the SPS-7 sample processed at 1000 °C for 5 minutes exhibits numerous small accumulated impurity regions, as depicted in Figure 3(j)-(l).

The accumulated area decreased further, and many small impurity areas were observed for the SPS-8, which was sintered at 1000 °C for 10 minutes, in comparison to the SPS-7 sample (Figure S2(p)-(r)). The increased temperature and prolonged sintering duration for SPS-7 and SPS-8 have reduced the region of accumulated impurity phases to smaller areas compared to the SPS-3 and SPS-4 samples. Microstructural analyses suggest that the optimal conditions for synthesizing F-doped SmFeAsO involve a temperature of 900 °C maintained for a duration of 5-10 minutes using the SPS process (table 2). This observation is supported by elemental mapping and aligns with previous reports [19] indicating that higher sintering temperatures and longer heating durations are unsuitable for F doping in Sm1111 system. The optimized samples (SPS-3 and SPS-4) exhibit some areas of accumulated impurities, while the remaining sample areas reveal a clean superconducting phase with numerous well-connected grain boundaries, as shown in Figure S2(g-i) and Figure 3(g-i). As observed for the parent P (Figure 3(a)–(c)), impurity phases appear throughout the entire region of the bulk when sintering conditions deviate from the optimal parameters (900°C for 5–10 minutes) employed in the SPS process. The findings from these observations are consistent with reports from the HP-HTS procedure [19].

## 4. Transport properties

Figure 4(a) depicts the temperature dependence of the resistivity for different SPS bulks, including both parent and HIP samples across the temperature range of 7 to 300 K. The parent sample exhibits a room temperature resistivity of approximately 6.9 mΩ-cm, which decreases linearly with decreasing temperature and a superconducting transition is observed below 55 K (Figure 4(a)). This behavior resembles that of previously reported F-doped Sm1111 [19, 33]. Furthermore, the HIP sample produced by the HP-HTS process also exhibits the same behavior as the parent P, as illustrated in the inset of Figure 4(a). The resistivity of the HIP sample is approximately 50% lower than that of the parent sample P, which is probably attributed to an 8% increase in the sample density (Table 3). The samples SPS-1 and SPS-2, processed at 600 °C for 5 and 10 minutes, respectively, exhibit the resistivity values that are threefold higher than that of the parent sample (refer to the inset Figure 4(a)) and they do not achieve zero resistance. However, their resistivity behaviors are comparable to those of HIP and P samples. The significant increase in the resistivity value could be attributed to the low sample density (42-43%) and the inhomogeneities presented in the constituent elements, as indicated by their elemental mappings. The resistivity of the four bulks, SPS-3 to SPS-6, which were prepared at 900 °C for 5, 10, 20, and 30 minutes, was noticeably lower than that of the parent P. The reduced resistivity observed in SPS-3 could be due to the result from the extensive homogeneous superconducting area attributed to the accumulation of impurity phases and the significant increase in sample density relative to the parent and HIP samples, as previously discussed in the microstructural analysis. In SPS-4, where the sintering time is extended to 10 minutes, the resistivity value exhibits a slight increase throughout the entire temperature range in comparison to SPS-3. However, both samples maintain almost the same sample density (Table 3). The increase in the resistivity may result from a minor alteration in the homogeneity of elemental distribution within the samples. The normal state resistivity of SPS-5 and SPS-6 increases with extended sintering times (20-30 min) across the entire temperature range. It may be attributed to a reduction in the size of accumulated impurity phases and the presence of observed many impurity phases in multiple locations of the sample, in contrast to SPS-3 and SPS-4. All SPS samples sintered at 900 °C exhibit almost the same sample density (Table 3) and resistivity characteristics. The variation in the observed resistivity values can be attributed to the distribution of impurity phases within the samples, corroborating the earlier discussion with microstructural analysis. Increasing the sintering temperature to 1000 °C, specifically for SPS-7 and SPS-8 for durations of 5 and 10 minutes, results in comparable resistivity behavior, though with a slightly reduced resistivity in comparison to SPS-5 and SPS-6. The observed inhomogeneity in these samples, attributed to the presence of impurity phases, suggests that the marginally lower resistivity can be ascribed to the increased compactness and a slight enhancement in sample density (approximately 1%), as indicated in Table 3. Furthermore, the resistivity behavior of these SPS samples closely resembles that of the other samples.

To analyse the superconducting transition of these samples, the low-temperature resistivity behaviors are depicted in Figure 4(b). This figure covers the temperature range from 30 K to 60 K and includes a comparison with the parent P and HIP bulks. The parent sample exhibited an onset transition temperature of 53.7 K and an offset transition temperature of 40 K. The HIP sample demonstrates a marginally lower onset transition temperature ($T_c^{onset}$) of 53 K, with an offset $T_c$ ($T_c^{offset}$) of 42 K. The reduced transition broadening is attributed to the improved sample density and grain connections. SPS-1 and SPS-2 samples have not reached zero resistivity until the measured temperature up to 7 K, suggesting the existence of a non-superconducting phase and weak link behaviors attributed to the presence of cracks and pores (Figure S2). The consistent $T_c^{onset}$ indicates that there has been no change in the fluorine content compared to the parent P, as confirmed by the structural analysis. All SPS bulks prepared at 900°C, specifically SPS-3, SPS-4, SPS-5, and SPS-6, exhibit a $T_c^{onset}$ of ~53-53.5 K, comparable to the parent sample, but with a slightly sharper transition width than the parent P and HIP samples. Increasing the sintering temperature to 1000 °C for 5-10 minutes, as observed for SPS-7 and SPS-8, results a decrease in $T_c^{onset}$ by ~2 K, suggesting a slight reduction in the actual content of fluorine within the superconducting 1111 lattice, supporting the above-discussed analysis. The reduction in the broadening of the superconducting transition for the samples prepared at 900 °C and 1000 °C via SPS indicates the improved grain connectivity due to the increased sample density. The analysis of the resistivity measurements reveals that the SPS process at 900°C for a duration of 5-10 minutes is the most effective for improving the sample quality and superconducting transition temperature of F-doped Sm1111 bulks compared to other SPS conditions.

## 5. Magnetic properties:

We have performed DC magnetic susceptibility measurements to confirm the superconductivity of these SPS bulks. Measurements were performed in both zero-field-cooled (ZFC) and field-cooled (FC) modes across a temperature range of 10-60 K, employing an applied magnetic field of 20 Oe for the selected samples. The magnetization curve was normalized using the magnetization value at 10 K to enable a comparative analysis of these bulk samples, as illustrated in Figure 5(a). The parent sample exhibits a transition temperature of 52.9 K and the observed negative value for the FC magnetization suggests considerable vortex pinning in the bulk samples, implying the existence of permanently trapped magnetic flux within the sample [19]. All SPS samples demonstrate marginally large negative FC values compared to the parent and HIP samples. The HIP sample exhibits an onset $T_c$ comparable to that of the parent P. The samples prepared at 900 °C, specifically SPS-4 and SPS-5, exhibit an onset $T_c$ of approximately 53 K, which closely aligns with the resistivity measurements. The long-sintered sample, SPS-6, exhibits a transition temperature of 52.5 K, which is 0.5 K lower than that of SPS-4, suggesting a minor evaporation of fluorine from this sample during SPS process. The samples SPS-7 and SPS-8 exhibit a transition temperature of 51.2 K, which is 1.7 K lower than that of the parent P, as indicated also by the resistivity measurements. The parent P and HIP samples show two-step transitions, which suggest poor intergrain connections, i.e., a weak-link behaviour as reported for FBS [19,33]. All SPS bulks have demonstrated a one-step transition, indicating that these samples exhibit better grain connections through the SPS process in comparison to the parent P and HIP. This facilitates the improvement of sample density, as presented in Table 3. Hence, these magnetic studies suggest that the SPS process at 900 °C for 5-10 minutes is adequate for the preparation of F-doped SmFeAsO bulks.

The critical current density ($J_c$) is a significant parameter for the practical applications of a superconductor. The magnetic hysteresis (*M-H*) loop was measured at 5 K for both the parent and HIP samples, as well as for the selected SPS samples prepared using the SPS technique at 45 MPa under a magnetic field up to 9 T. The hysteresis loop for the parent, HIP, and SPS-7 samples is illustrated in the inset of Figure 5(b) and demonstrates the behavior consistent with previous studies [32,33]. The width of the hysteresis loop (*Δm*) is determined by the analyzing the magnetic moment in both ascending and descending magnetic fields. The Bean model [38] can be used to determine the critical current density using the formula: $J_c = 20\Delta m/Va(1-a/3b)$, where *a* and *b* represent the sample dimensions, *V* is the sample's volume. The rectangular shaped samples were employed for magnetic measurements [38]. Considering the sample dimensions, the calculated critical current density is illustrated in Figure 5(b). The parent compound exhibits a $J_c$ value of $10^3$ A/cm² at 0.5 T, which decreases slightly with the applied magnetic field up to 9 T. This behavior is similar to that reported for F-doped Sm1111 [6]. The $J_c$ value for the HIP sample is improved within the entire magnetic field range, and reaches to the value of $3 \times 10^3$ A/cm² at 0.5 T. There is a slight increase in the $J_c$ value for the SPS samples compared to the parent P but approximately the same as that of the HIP sample. This slight improvement in $J_c$ value may be attributed to the enhanced sample density or homogeneity in the sample by SPS process, as evidenced by the structural and microstructural analysis. SPS samples have almost two times higher sample density than parent P, however, we could not observe a very high $J_c$, which could be due to the presence of impurity phases that neutralize its effect. To analyse the pinning force characteristics of these samples, we employed the correlation between the applied magnetic field and the density of vortex pinning force, $F_p$, utilizing the equation $F_p = \mu_0 H \times J_c$ [39]. Figure S3 in the supplementary data file illustrates the calculated pinning force for the samples: parent, HIP, SPS-4, and SPS-7. The enhancement of the critical current density ($J_c$) is attributed to an increase in the effective pinning force ($F_p$), which is linked to microstructural modifications such as enhanced grain connectivity, greater density, and refined porosity. This finding aligns with previous reports based on other families of FBS [19].

## 6. Discussion:

To summarize the findings from the Sm1111 bulks synthesized by the SPS technique, we have plotted the onset transition temperature ($T_c^{onset}$), transition width ($\Delta T$), room temperature resistivity ($\rho_{300K}$), Residual Resistivity Ratio ($RRR = \rho_{300K} / \rho_{60K}$), and critical current density ($J_c$), including both the parent and HIP samples in Figure 6. The $T_c^{onset}$ is 53.7 K for the parent sample, while the HIP sample shows a slightly lower onset transition temperature of 53 K. The SPS samples, from SPS-1 to SPS-4, display nearly the same transition temperatures as that of the parent sample, as shown in Figure 6(a). However, $T_c^{onset}$ starts to decrease for SPS-5 and SPS-6, specifically during the long sintering duration at 900 °C. Furthermore, for SPS-7 and SPS-8 samples sintered at 1000 °C, $T_c^{onset}$ decreases rapidly, which is expected due to a small reduction of fluorine content within the Sm1111 lattice, as discussed above with XRD data analysis. Figure 6(b) shows the broadening of the transition width ($\Delta T$) for the various bulks, which is determined by a subtraction of the $T_c^{onset}$ and the offset transition temperature ($T_c^{offset}$). The parent P has a transition broadening of around 13.5 K, which is reduced to 12 for HIP sample. The SPS-1 and SPS-2 samples have not reached to zero resistance (i.e., no $T_c^{offset}$), so there are no respective points for these samples in Figure 6(b). The $\Delta T$ reduces again for SPS-3 and SPS-4, attaining a minimum for the SPS-4 bulk. This transition width increases almost linearly for SPS-5 to SPS-6 samples, after which it remains relatively constant. The variation of room temperature resistivity ($\rho_{300K}$) for different samples is depicted in Figure 6(c). The resistivity $\rho_{300K}$ value is 6.97 milliohm-cm for the parent sample, which exhibits a modest reduction for HIP-1 attributed to a minor enhancement in the sample density (Table 3). However, this value is increased rapidly for the samples SPS-1 and SPS-2, which could be due to the low sample density and weak connections between grains, as observed from structural and microstructural analysis. The SPS process conducted at 900 °C reduces the resistivity value rapidly to 0.78-0.85 milliohm-cm with the sintering time of 5 and 10 min for the SPS-3 and SPS-4 bulks, respectively. For SPS-5 and SPS-6 samples, $\rho_{300K}$ value has slightly increased to ~1.2 milliohm-cm, supporting the existence of the accumulated impurity areas. The SPS-7 and SPS-8 samples exhibit a slight reduction in the $\rho_{300K}$ value to 0.9 milliohm-cm, potentially attributed to marginal improved in the intergrain connections (i.e., the sample density) (Table 3). Residual resistivity ratio (*RRR*) variation with different samples is shown in Figure 6(d). The parent P has an *RRR* value of 3.5, which is increased to 4 for the HIP sample. SPS-1 and SPS-2 have very low *RRR* value, supporting the inhomogeneity of these samples as observed from the microstructure and resistivity measurements. This value again increased to 4.47 for SPS-3 and 4.29 for SPS-4. Interestingly, other samples SPS-5 to SPS-8 have almost the same *RRR* value. It suggests that overall, these samples have almost the same homogeneity for either long sintering duration or high sintering temperature. Figure 6(e) depicts the variation of critical current density $J_c$ at the magnetic field of 0.2 T and 8 T for the samples prepared by CSP, HP-HTS, and SPS processes. The $J_c$ value is increased for HIP samples compared to parent P in the whole magnetic fields. SPS samples have shown an increment of $J_c$ value continuously from SPS-3 to SPS-7 samples, while SPS-8 exhibits a value nearly identical to that of SPS-7. It is noteworthy that the order of $J_c$ value is almost the same for all SPS and HIP bulks, however, there is a slight improvement compared to the parent P. This small enhancement of $J_c$ could be due to slightly improved pinning force as shown in the supplementary Figure S3. Basically, these analyses confirm that the optimal conditions for the SPS process for F-doped Sm1111 are a temperature of 900 °C maintained for a duration of 5-10 min, while other synthesis conditions reduce the sample quality and superconducting properties of Sm1111 bulks.

In Figure 7, we have prepared a comparative graph, depicting the performance of different families of iron-based superconductors, specifically K-doped $BaFe_2As_2$ (Ba122) [27], [40] and $CaKFe_4As_4$ (Ca1144) [28, 41] alongside our results for F-doped Sm1111 processed by SPS. These Ba122 and Ca1144 superconductors were prepared by the CSP method at ambient pressure and the SPS technique at ~45-50 MPa under optimum conditions. The details about the superconducting parameters, sample density, and the impurity phase are mentioned in Table 4 in the supplementary data file. Ba122 bulk has the sample density of 68% by the CSP process which is increased to 96% by the SPS process. The SPS process of Ba122 has reduced the impurity phase, and the $J_c$ value is enhanced by one order of magnitude compared to CSP process. In the case of Ca1144, the SPS process also improved the sample density from 66% to 96%, and the impurity phases are reduced compared to the CSP process. Hence, the critical current density is enhanced by one order of magnitude compared to that of CSP, as shown in Figure 7. Interestingly, the SPS method reduces the impurity phases and increases the sample density, resulting a reported boost of $J_c$ for Ba122 [27] and Ca1144 [28]. For F-doped Sm1111, the sample density is increased almost double compared to the CSP process, i.e., from 50 to 98%; however, the amount and type of impurity remain consistent across all SPS and CSP processed F-doped Sm1111. Probably, the presence of the impurities could be a reason for no significant enhancement of $J_c$ for F-doped SmFeAsO prepared by SPS. In this bulk sample, the 1111 phase contains the impurities such as SmOF and SmAs phases, which remain nearly constant after employing the SPS technique. Interestingly, the amount of these impurity phases remains unchanged despite high-pressure growth, and a slight improvement in $J_c$ has been observed, as recently reported elsewhere [19] which is agreed well with the findings from our SPS bulks. The SPS process demonstrates significant effectiveness in enhancing sample density, which is a crucial factor for compacting the sample and improving intergrain connections, particularly from the application perspective for various FBS families. The processing of Ba122 and Ca1144 bulks by SPS resulted in a reduction of impurity phases, leading to an observed enhancement of $J_c$ alongside the improved sample density. Nonetheless, for F-doped Sm1111 bulks produced through SPS, there exists an issue with the presence of impurity phases that closely resemble those found in the CSP process. As a result, a marginal enhancement or nearly identical value of $J_c$ is observed for the SPS-processed SmFeAsO$_{0.80}$F$_{0.20}$ bulks.

Generally, there are two types of critical current densities found in a polycrystalline FBS material: intragrain $J_c$ and intergrain $J_c$. Previous studies [13, 42-44] suggest that superconductivity within each grain is sufficiently robust; however, the bulk critical current is limited by intergrain currents that flow across the grain boundaries in a polycrystalline sample. The intergrain coupling at these grain boundaries plays a crucial role in evaluating their transport applicability. In FBS, grain boundaries are weakly connected due to their intrinsic weak-link behavior and/or extrinsic structural factors, which leads to a disparity of two to three orders of magnitude between intergrain and intragrain current densities. The $J_c$, obtained from a magnetic hysteresis loop measurement, presents an average value that reflects both intergrain and intragrain current densities. Various reports [13, 42-44] suggest that non-superconducting phases occupy at least three-quarters of the Sm1111 grain boundaries, resulting in a current-blocking network. This phenomenon may account for the significant difference between the measured intergrain and the intragrain $J_c$, as well as decrease the overall $J_c$ value of this Sm1111 superconductor. The presence of impurity phases extrinsically limits the intergranular $J_c$ on a macroscale. Thus, the variation in the intergrain $J_c$ among different bulks may stem from macroscopic inhomogeneity and impurity phases at the grain boundaries. The reduced intergrain connectivity can be explained by considering the formation of weak Josephson junctions that arise from the presence of non-superconducting phases, which tend to increase as the amount of non-superconducting phase increases [13, 42-44]. On the basis of these studies, we propose that the weak intergrain coupling caused by impurity phases between grains in these SPS-processed F-doped Sm1111 could be a contributing factor to the lack of significant improvement in $J_c$. Further investigations are required in this area utilizing the SPS technique to minimize the presence of impurity phases and consequently, a significant improvement in the superconducting characteristics of F-doped Sm1111 can be anticipated, akin to the observations made in the Ba122 and Ca1144 families.

## Conclusions

SmFeAsO$_{0.80}$F$_{0.20}$ bulks were prepared using the spark plasma sintering (SPS) method by optimizing the sintering temperature (600-1000°C) and duration (5-30 min). A broad characterization has been conducted for all these bulks to elucidate the impact of the SPS process on their superconducting properties. F-doped Sm1111 bulks were also synthesized using the CSP and HP-HTS methods at 900°C to make a comprehensive understanding of these three synthesis processes in this study. CSP method depicted the superconducting transition of 53-54 K, a sample density of 50%, and the critical current density of $10^3$ A/cm², whereas HP-HTS process exhibited a slight increase in the sample density (58%) and critical current density ($3 \times 10^3$ A/cm²), with an unchanged $T_c$ of 53-54 K. Our studies employing SPS showed that a sintering temperature below 900°C is insufficient to compact the sample or to produce the homogenous Sm1111 bulks. On the other hand, the elevated sintering temperature (1000 °C) or prolonged sintering duration (>10 min) reduces the superconducting transition, possibly due to the evaporation of lighter elements such as fluorine and arsenic during the SPS process. However, Sm1111 prepared at 900°C for a sintering duration of 5-10 min is sufficient to produce good quality samples compared to other SPS conditions. Interestingly, SPS has improved the sample density by almost two times (~97-98%), although the type and quantity of impurity phases remain consistent across all samples, as also observed in the CSP and HP-HTS processes. A comparative investigation of the CSP, HP-HTS, and SPS techniques suggests that none of them effectively reduce the impurity phases (SmAs and Sm$_2$O$_3$/SmOF) during the synthesis of the SmFeAs(O,F) phase, resulting in no significant enhancement of the superconducting properties. These results differ from those reported for other FBS families (122 and 1144). Our studies suggest that high-pressure synthesis and the SPS approach should be applied only on pure phase F-doped SmFeAsO bulks; thus, the enhancement of superconducting properties can be expected due to the improved sample density. These findings will contribute significantly to the advancement of FBS, particularly in the development of superconducting wires and tapes, highlighting the need for further research in this area.

**Acknowledgments:**

The work was funded by SONATA-BIS 11 project (Registration number: 2021/42/E/ST5/00262) sponsored by National Science Centre (NCN), Poland. SJS acknowledges financial support from National Science Centre (NCN), Poland through research Project number: 2021/42/E/ST5/00262.

# References

[1] Kamihara Y, Watanabe T, Hirano M, Hosono H. Iron-based layered superconductor La[O$_{1-x}$F$_x$]FeAs (x = 0.05-0.12) with Tc = 26 K. J. Am. Chem. Soc 2008;130:3296-329.

[2] Ren ZA, Zhao ZX. Research and prospects of iron-based superconductors. Adv. Mater 2000;21:4584-4592.

[3] Paglione J, Greene RL. High-temperature superconductivity in iron-based materials. Nature Physics 2010;6:645-658.

[4] Shimoyama JI. Potentials of iron-based superconductors for practical future materials. Supercond. Sci. Tech. 2014;27:044002.

[5] Singh SJ. Sturza MI. Bulk and single crystal growth progress of iron-based superconductors (FBS): 1111 and 1144. Crystal 2022;12:20.

[6] Singh SJ, Shimoyama J, Yamamoto A, Ogino H, Kishio K. Transition temperature and upper critical field in SmFeAsO$_{1-x}$F$_x$ synthesized at low heating temperatures. IEEE Transactions on Applied Superconductivity 2013;23:2239352.

[7] Fujioka M, Denholme SJ, Ozaki T, Okazaki H, Deguchi K, Demura S, Hara H, Watanabe T, Takeya H, Yamaguchi T, Kumakura H, Takano Y. Phase diagram and superconductivity at 58.1 K in α-FeAs-free SmFeAsO$_{1-x}$F$_x$. Supercond. Sci. Tech. 2013; 26:085023.

[8] Qi Y, Gao Z, Wang L, Wang D, Zhang X, Ma Y. Superconductivity in co-doped SmFeAsO', Supercond Sci Technol 2008;21:115016.

[9] Singh SJ, Shimoyama J, Yamamoto A, Ogino A, Kishio K. Effects of Mn and Ni doping on the superconductivity of SmFeAs(O,F). Physica C 2013;494:57-61.

[10] Singh SJ, Shimoyama SJ, Yamamoto A, Ogino H, Kishio K. Effects of phosphorus doping on the superconducting properties of SmFeAs(O,F). Physica C 2016;504:1923.

[11] Yong LC, Cheng CH, Cui YJ, Zhang H, Zhang Y, Yang Y, Zhao Y. Ir doping-induced superconductivity in the SmFeAsO system. Journal of the American Chemical Society 2009:31,10338-10339.

[12] Wang XC, Yu J, Ruan B, Pan B, Mu Q, Liu T, Zhao K, Chen G, Ren Z. Revisiting the Electron-Doped SmFeAsO: Enhanced Superconductivity up to 58.6 K by Th and F Codoping', Chinese Physics Letters 2017;34:077401.

[13] Singh SJ, Shimoyama J, Yamamoto A, Ogino H, Kishio K. Significant enhancement of the intergrain coupling in lightly F-doped SmFeAsO superconductors. Supercond. Sci. Tech. 2013;26:065006.

[14] Hosono H, Kuroki K. Iron-based superconductors: Current status of materials and pairing mechanism. Physica C 2015;514:399-422.

[15] Singh P, Manasa M, Azam M, Singh SJ. High-pressure growth effect on the properties of high-Tc iron-based superconductors: A short review. Cryogenics 2025;147:104028.

[16] An RZ, Wei L, Jie Y, Wei Y, Xiao-Li S, Cai Z, Can C, Li DX, Ling S, Fang Z, Xian Z. Superconductivity at 55 K in Iron-Based F-Doped Layered Quaternary Compound Sm[$O_{1-x}F_x$]FeAs. Chin. Phys. Lett. 2008;25:2215.

[17] Ren Z, Yang J, Lu W, Yi W, Shen X, Li Z, Che G, Dong X, Sun L, Zhou F. Superconductivity in the iron-based F-doped layered quaternary compound Nd[$O_{1-x}F_x$]FeAs. EPL 2008;82:57002.

[18] Azam M, Manasa M, Morawski A, Cetner T, Singh SJ. High Gas Pressure and High-Temperature Synthesis (HP-HTS) Technique and Its Impact on Iron-Based Superconductors. Crystals 2023;13:1525.

[19] Azam M, Manasa M, Zajarniuk T, Diduszko R, Palasyuk T, Cetner T, Morawski A, Jastrzębski C, Szewczyk A, Wierzbicki M, Singh SJ. High-pressure growth effects on the superconducting properties of Sm-based oxypnictide superconductors. Ceramics International 2025;51: 13734-13751.

[20] Karpinski J, Zhigadlo ND, Katrych S, Bukowski Z, Moll P, Weyeneth S, Keller H, Puzniak R, Tortello, Daghero D, Gonnelli R, Aprile IM, Fasano Y, Fischer, Rogacki K, Batlogg B. Single crystals of LnFeAs$O_{1-x}F_x$ (Ln = La, Pr, Nd, Sm, Gd) and Ba$_{1-x}$Rb$_x$Fe$_2$As$_2$: Growth structure and superconducting properties. Physica C 2009;469:370-380.

[21] Zhigadlo ND, Weyeneth S, Katrych S, Moll PJ, Rogacki K, Bosma S, Puzniak R, Karpinski J, Batlogg B. High-pressure flux growth, structural, and superconducting properties of LnFeAsO (Ln = Pr, Nd, Sm) single crystals. Phys Rev B 2012;86:214509, 2012.

[22] Olevsky EASpark-Plasma Sintering and Related Field-Assisted Powder Consolidation Technologies. Materials 2017. https://doi.org/10.3390/books978-3-03842-383-6

[23] Kumar DB, Babu BS, Aravind Jerrin KM, Joseph N, Jiss A. Review of spark plasma sintering process. IOP Conf. Ser : Mater. Sci. Eng 2020;993:012004.

[24] Oliver UC, Sunday AV, Ikechukwu Christain EI, Elizabeth MM. Spark plasma sintering of aluminium composites-a review. Int J Adv Manuf Technol 2021;112:1819-1839.

[25] Tokita M. Progress of spark plasma sintering (SPS) method, systems, ceramics applications and industrialization. Ceramics 2021; 4: 160-198.

[26] Zaikina JV, Kwong MY, Baccam B, Kauzlarich SM. Superconductor-in-an-Hour: Spark Plasma Synthesis of Co- and Ni-Doped BaFe$_2$As$_2$. Chemistry of Materials 2018;30: 8883-8890.

[27] Tokuta S, Hasegawa Y, Shimada Y, Yamamoto A. Enhanced critical current density in K-doped Ba122 polycrystalline bulk superconductors via fast densification. iScience 2022;25:03992.

[28] Ishida S, Naik SP, Tsuchiya Y, Mawatari Y, Yoshida Y, Iyo A, Eisaki H, Kamiya Y, Kawashima H, Ogino H. Synthesis of CaKFe$_4$As$_4$ bulk samples with high critical current density using a spark plasma sintering technique. Supercond. Sci. Technol. 2020; 33:094005.

[29] Puneet P, Podila R, He J, Rao M, Howard A, Cornell N, Zakhidov AA. Synthesis and superconductivity in spark plasma sintered pristine and graphene-doped FeSe$_{0.5}$Te$_{0.5}$', Nanotechnol Rev 2015;4:411-417.

[30] Kursumovic A, Durrell JH, Chen SK, MacManus-Driscoll JL. Ambient/low pressure synthesis and fast densification to achieve 55K $T$c superconductivity in NdFeAsO$_{0.75}$F$_{0.25}$. Supercond. Scien. Technol. 2010;23:025022.

[31] Ishida S, Kamiya Y, Tsuchiya Y, Naik PK, Mawatari Y, Iyo A, Yoshida Y, Eisaki H, Kawashima K, Ogino H. Synthesis of c-axis textured CaKFe$_4$As$_4$ superconducting bulk via spark plasma texturing technique. Journal of Allys and Compounds 2023;961:171093.

[32] Azam M, Manasa M, Zajarniuk T, Diduszko R, Cetner T, Morawski A, Singh SJ. Antimony Doping Effect on the Superconducting Properties of SmFeAs(O,F). IEEE Transactions on Applied Superconductivity 2024;34:3343328.

[33] Azam M, Manasa M, Zajarniuk T, Palasyuk T, Diduszko R, Cetner T, Morawski A, Jastrzebski C, Wierzbicki M, Wiśniewski A, S. J. Singh, "Copper doping effects on the superconducting properties of Sm-based oxypnictides, "J. Am. Ceram. Soc. 2024;107: 6806-6820.

[34] Iebole M, Braccini V, et al. Fe(Se,Te) Thin Films Deposited through Pulsed Laser Ablation from Spark Plasma Sintered Targets. Materials 2024, 2594.

[35] Johnston DC. The puzzle of high temperature superconductivity in layered iron pnictides and chalcogenides. Adv Phys. 2010;59:803-61.

[36] Ingle F, Priolkar KR, Pal A, Awana A, Emura S. Local structural distortions and their role in Superconductivity in SmFeAsO$_{1-x}$F$_x$ superconductors. Supercond. Sci. Technol. 2014; 27:075010

[37] Fujioka M, Denholme SJ, Tanaka M, Takeya H, Yamaguchi T, Takano Y. The effect of exceptionally high fluorine doping on the anisotropy of single crystalline SmFeAsO$_{1-x}$F$_x$', Appl. Phys. Lett. 2014;105:10260.

[38] Bean CP. Magnetization of high-field superconductors. Rev. Mod. Phys.1985;36:31.

[39] Dew-Hughes D. Flux pinning mechanisms in type II superconductors. Philosophical Magazine 1974;30:293-305. doi: 10.1080/14786439808206556.

[40] Weiss JD, Jiang J, Polyanskii AA, Hellstrom EE, Mechanochemical synthesis of pnictide compounds and superconducting Ba$_{0.6}$K$_{0.4}$Fe$_2$As$_2$ bulks with high critical current density. Supercond Sci Technol 2013;26:074003.

[41] Manasa M, Azam M, Zajarniuk T, R. Diduszko, Cetner T, Morawski A, Wiśniewski A, Singh SJ. Enhancement of Superconducting Properties of Polycrystalline CaKFe4As4 by High-Pressure Growth. IEEE Transactions on Applied Superconductivity 2024;34, 345821.

[42] Yamamoto A, Jiang J, Tarantini C, Craig N, Polyanskii E, et al. Evidence for electromagnetic granularity in the polycrystalline iron-based superconductor. Appl. Phys. Lett. 2008;92:252501.

[43] Kametani F, Polyanskii AA, Yamamoto A, Jiang J, Hellstrom E, Gurevich A, Larbalestier DC, Ren ZA, Yang J, Dong XL, Lu W and Zhao ZX. Combined microstructural and magneto-optical study of current flow in polycrystalline forms of Nd and Sm Fe-oxypnictides. Supercond. Sci. Technol. 2009;22:015010.

[44] Iida K, Hänisch J. Yamamoto A. Grain boundary characteristics of Fe-based superconductors. Supercond. Sci. Technol. 2020;33:04300

**Table 1:** List of the synthesis conditions and sample code for SmFeAsO$_{0.80}$F$_{0.20}$ bulks prepared by CSP, HP-HTS and SPS processes.

| Sample Synthesis conditions | Sample code |
| --- | --- |
| First step: 900 °C, 45 h, ambient pressure | Parent (P) |
| First step: 900 °C, 45 h, ambient pressure<br>Second step: Sintered at 900 °C, 1 h, 500 MPa | HIP |
| First step: 900 °C, 45 h, ambient pressure<br>Second step: Sintered at 600 °C, 5 minutes, 45 MPa | SPS-1 |
| First step: 900 °C, 45 h, ambient pressure<br>Second step: Sintered at 600 °C, 10 minutes, 45 MPa | SPS-2 |
| First step: 900 °C, 45 h, ambient pressure<br>Second step: Sintered at 900 °C, 5 minutes, 45 MPa | SPS-3 |
| First step: 900 °C, 45 h, ambient pressure<br>Second step: Sintered at 900 °C, 10 minutes, 45 MPa | SPS-4 |
| First step: 900 °C, 45 h, ambient pressure<br>Second step: Sintered at 900 °C, 20 minutes, 45 MPa | SPS-5 |
| First step: 900 °C, 45 h, ambient pressure<br>Second step: Sintered at 900 °C, 30 minutes, 45 MPa | SPS-6 |
| First step: 1000 °C, 45 h, ambient pressure<br>Second step: Sintered at 1000 °C, 5 minutes, 45 MPa | SPS-7 |
| First step: 1000 °C, 45 h, ambient pressure<br>Second step: Sintered at 1000 °C, 10 minutes, 45 MPa | SPS-8 |

**Table 2:** List of the synthesis parameters for our best F-doped SmFeAsO prepared by conventional synthesis process at ambient pressure (CSP), high gas pressure and high temperature synthesis (HP-HTS), and spark plasma sintering (SPS) methods.

| Synthesis Method | Sample Code | Pressure | Heating/Cooling Rate | Sintering Temperature | Time | Atmosphere Control |
|---|---|---|---|---|---|---|
| CSP | Parent (P) | Ambient | 1.5 °C/min | 900 °C | 45 h | Vacuum-sealed quartz tube |
| HP-HTS | HIP | 500 MPa | 7.5 °C/min | 900 °C | 1 h | Sealed in Ta tube under Ar atmosphere |
| SPS | SPS-3 to SPS-4 | 45 MPa | ~50 °C/min | 900 °C | 5-10 min | Vacuum environment |

**Table 3:** A list of the sample code, the amount of impurity phases (SmOF and SmAs) obtained from the XRD measurements and the calculated sample density.

| Sample Code | SmOF (%) | SmAs (%) | Sample density (%) |
|---|---|---|---|
| Parent | ~7 | ~2 | ~50 |
| HIP | ~6 | ~2 | ~58 |
| SPS-1 | ~4 | ~2 | ~43 |
| SPS-2 | ~4 | ~1 | ~42 |
| SPS-3 | ~4 | ~3 | ~92 |
| SPS-4 | ~4 | ~3 | ~97 |
| SPS-5 | ~4-5 | ~3 | ~97 |
| SPS-6 | ~4 | ~2 | ~97 |
| SPS-7 | ~4-5 | ~3 | ~98 |
| SPS-8 | ~4 | ~2 | ~98 |

**Figure 1:** **(a)** Powder XRD pattern of SmFeAsO$_{0.80}$F$_{0.20}$ bulks prepared by CSP, HP-HTS and various SPS-1 to SPS-8 bulks prepared by spark plasma sintering (SPS) with various sintering temperatures and time. **(b)** An enlarged view of the main peak (102) position is depicted for different samples. The variation of **(c)** lattice parameter (*a*), **(d)** lattice parameter (*c*), and **(e)** unit cell volume (*V*) with the various synthesis pressures for all SmFeAsO$_{0.80}$F$_{0.20}$ samples.

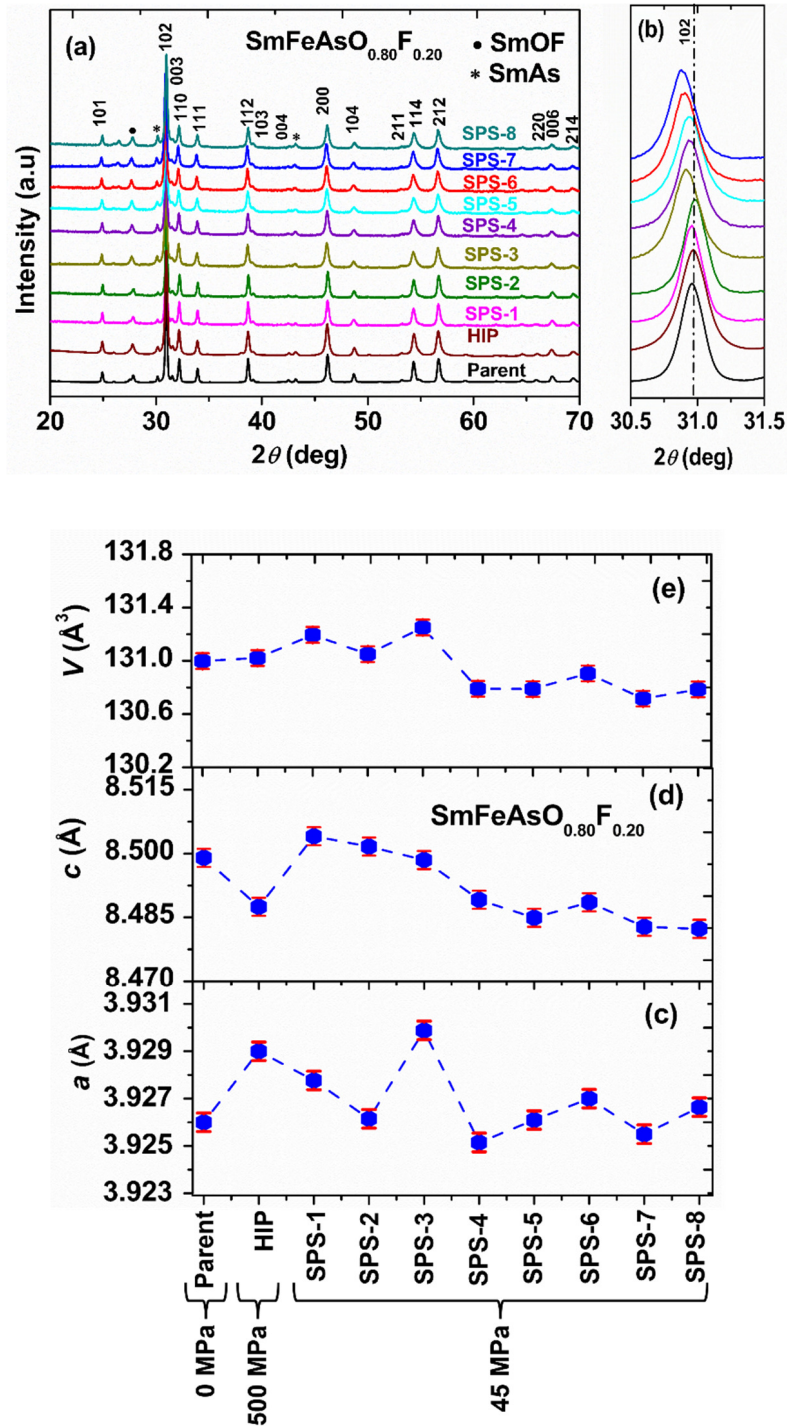

**Figure 2: (i)** Elemental mapping for all constituent elements of: **(i)** Parent P prepared by CSP, **(ii)** HIP prepared by HP-HTS, and **(iii)** SPS-4, **(iv)** SPS-7 prepared by SPS process. First and last images for each sample is SEM image and a combined image of all the constituent elements, respectively. The rest of the images depict the elemental mapping of an individual element Sm, Fe, As, O and F. The circles are used to depict the secondary phases, such as SmAs and SmOF ($Sm_2O_3$) or inhomogeneity of the constituent elements.

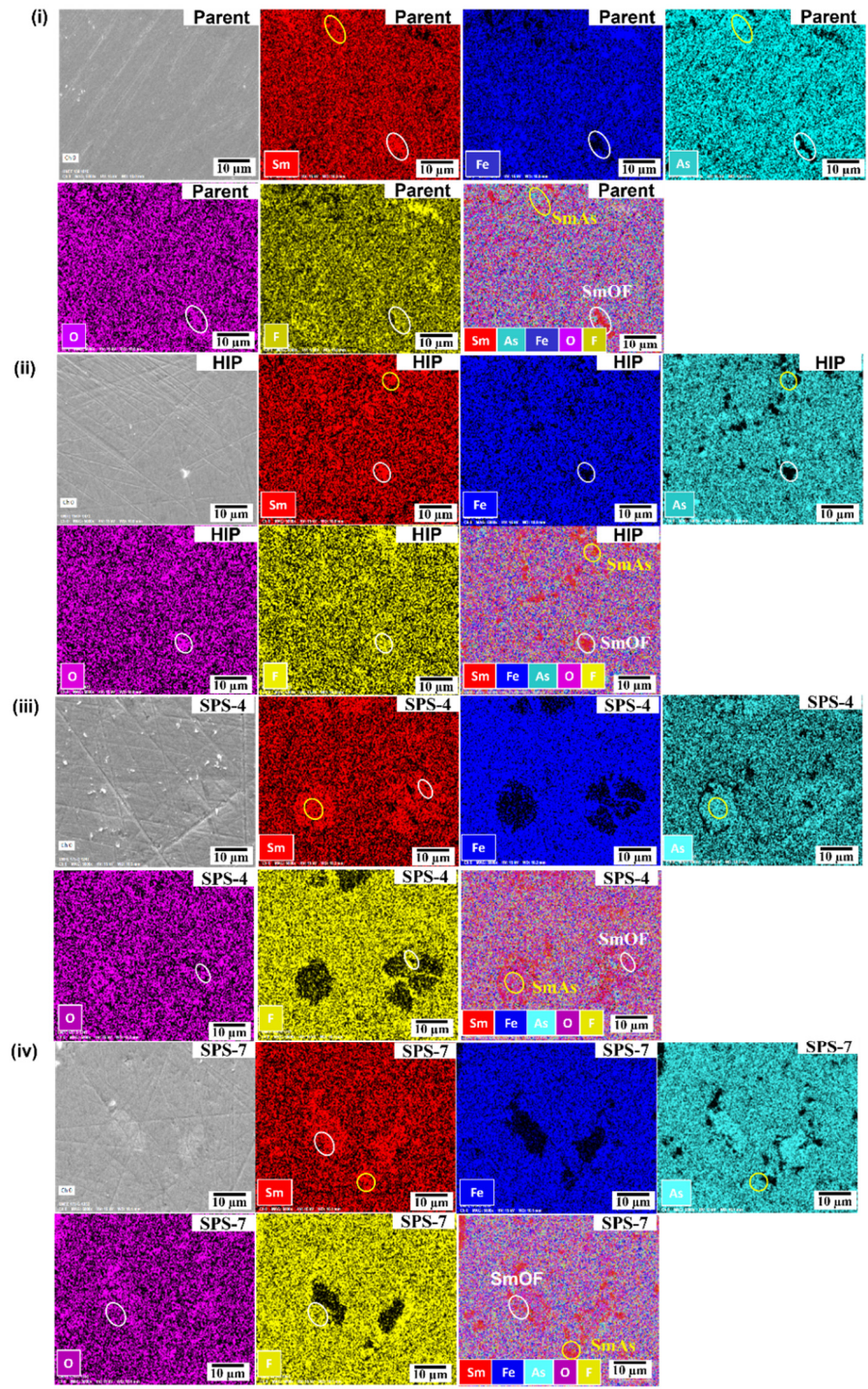

**Figure 3:** Backscattered electron images (BSE; AsB) of **(a)-(c)** Parent P, **(d)-(f)** HIP, **(g)-(i)** SPS-4, and **(j)-(l)** SPS-7. Bright contrast, light gray, and black contrast correspond to $Sm_2O_3$ (SmOF), $SmFeAsO_{0.80}F_{0.20}$ phase and pores, respectively. One can note that black contrast can occasionally be SmAs.

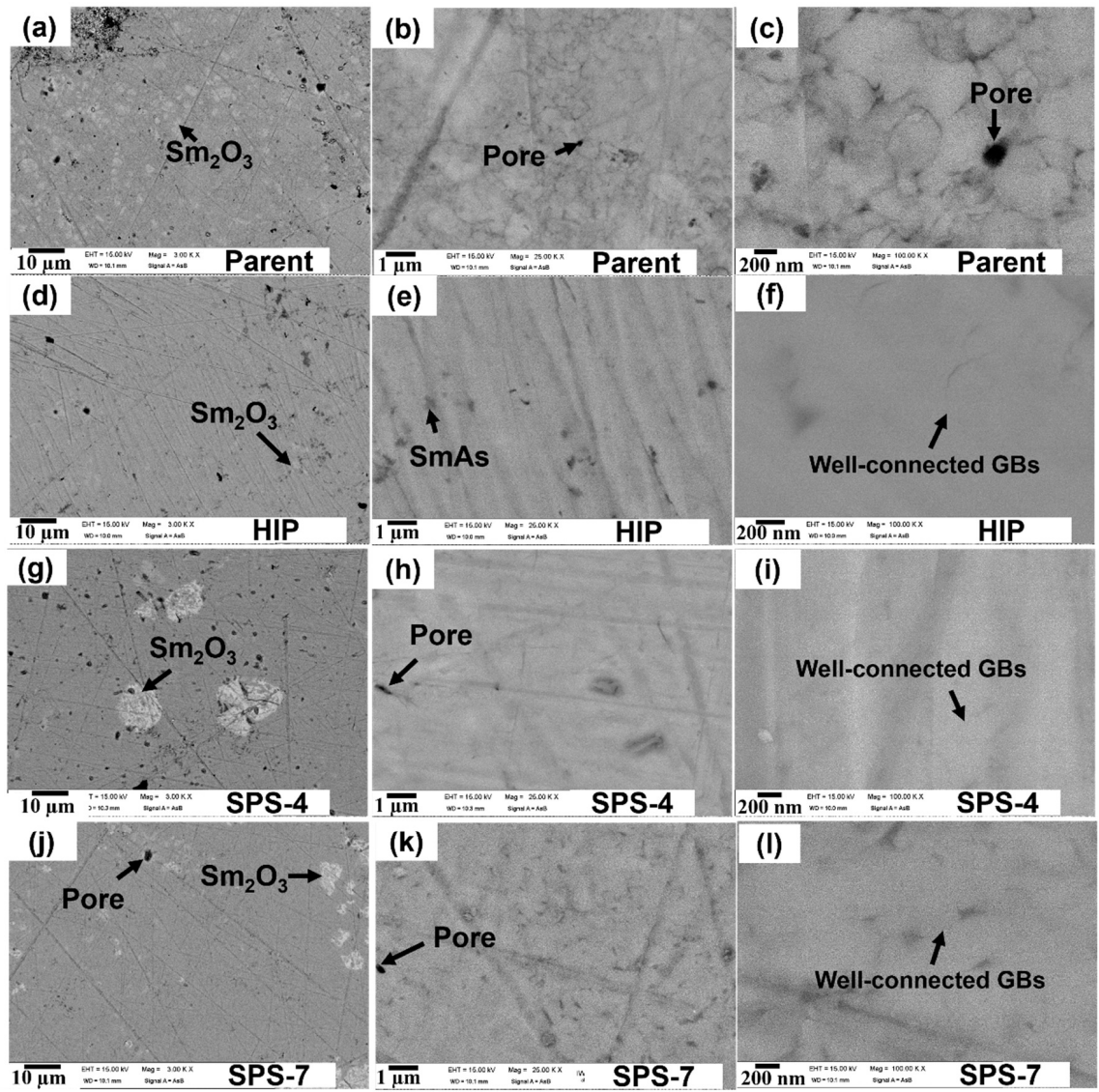

**Figure 4:** **(a)** The temperature variation of resistivity ($\rho$) up to the room temperature for various SPS bulks. The inset image depicts the temperature dependence of the resistivity for the parent, HIP, SPS-1 and SPS-2 up to room temperature. **(b)** Low-temperature variation of the resistivity up to 60 K for various SPS samples. The inset figure shows the low-temperature dependence of the resistivity for the parent, HIP, SPS-1, and SPS-2.

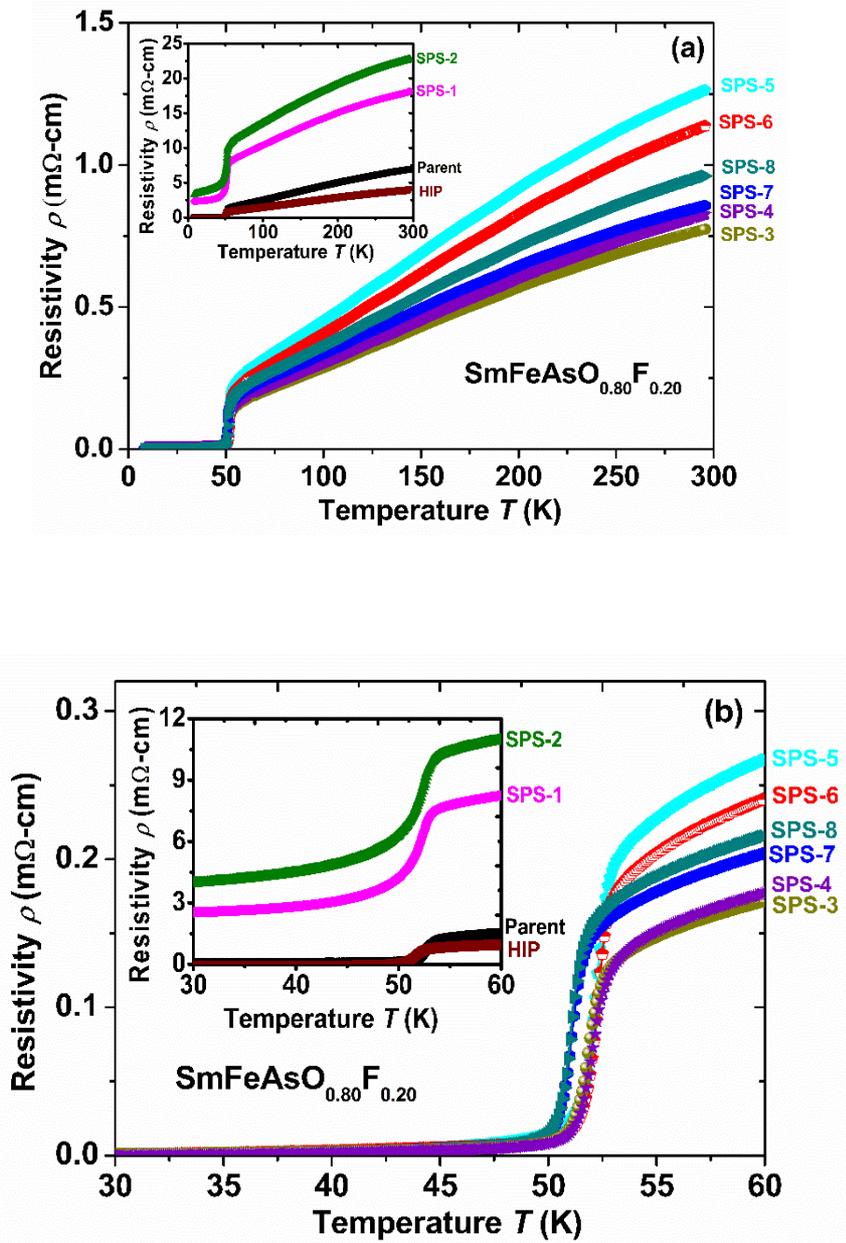

**Figure 5:** **(a)** The temperature dependence of the normalized magnetic moment in ZFC and FC modes under a magnetic field of 20 Oe. **(b)** The variation of the critical current density ($J_c$) at a temperature of 5 K with the applied magnetic field up to 9 T for the parent, HIP, SPS-4, SPS-5, SPS-6, SPS-7 and SPS-8 bulks. The inset of the figure (b) illustrates the magnetic hysteresis loop (*M–H*) for the parent, HIP and SPS-7 samples at a temperature of 5 K under the magnetic field up to 9 T.

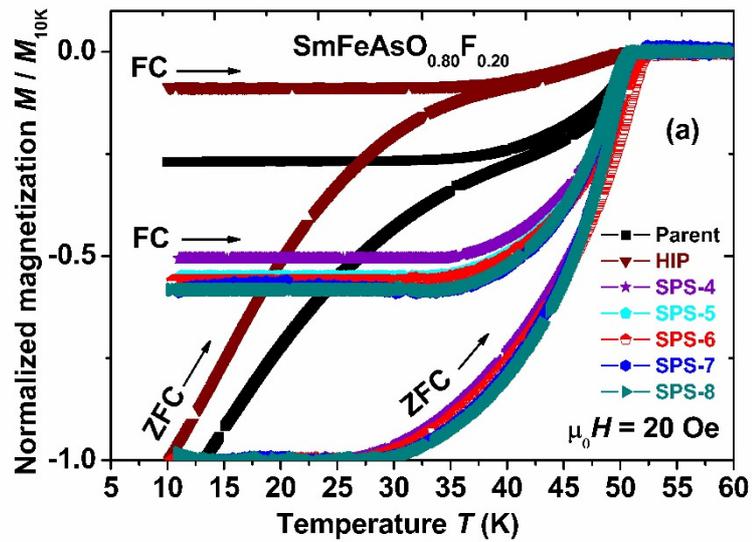

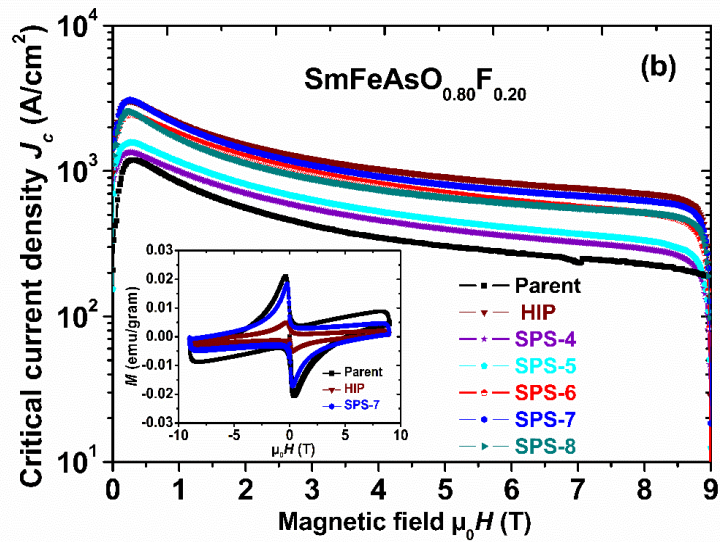

**Figure 6:** The variations of **(a)** the onset transition temperature ($T_c^{onset}$), **(b)** the transition width ($\Delta T$), **(c)** the room temperature resistivity ($\rho_{300\,K}$), **(d)** the Residual Resistance Ratio ($RRR = \rho_{300\,K} / \rho_{60\,K}$), and **(e)** the critical current density ($J_c$) of different SmFeAsO$_{0.80}$F$_{0.20}$ bulks prepared by CSP, HP-HTS and SPS processes under the different growth pressures.

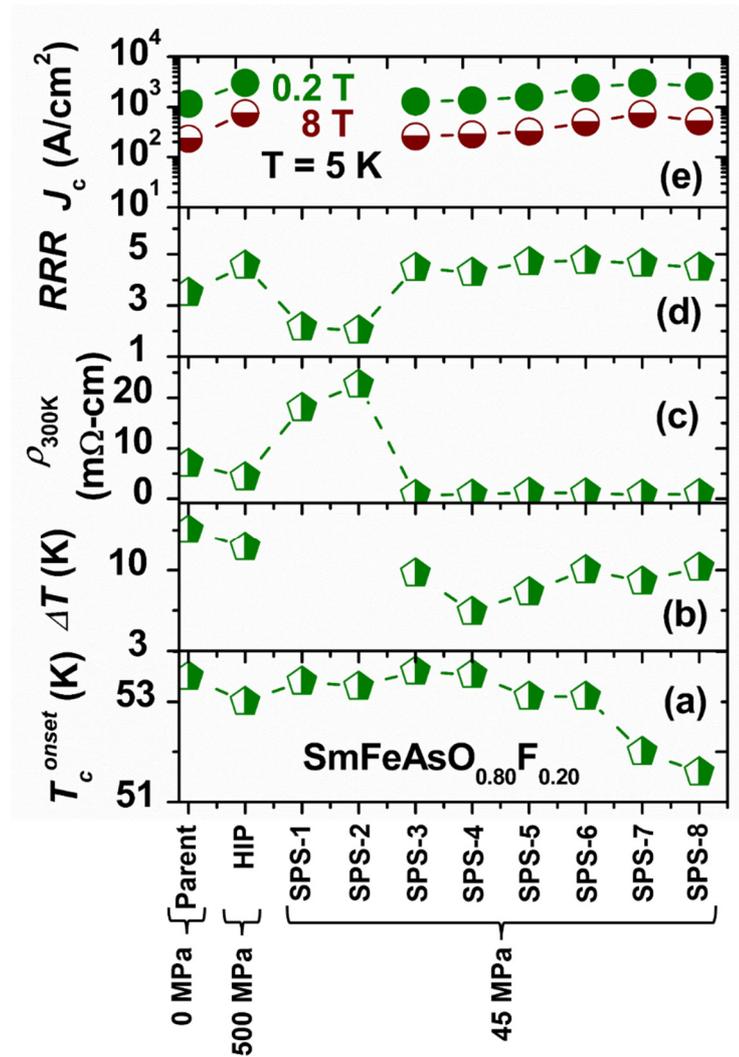

**Figure 7:** The magnetic field dependence of the critical current density ($J_c$) at ~5 K for our F-doped Sm1111 (Sm1111), (Ba,K)Fe$_2$As$_2$ (Ba122) [27], [40] and CaKFe$_4$As$_4$ (1144) [28], [41] bulks prepared by CSP process at ambient pressure (AP) and SPS method. The details about the $J_c$ and other parameters for these bulks are mentioned in Table 4 of the supplementary data file.

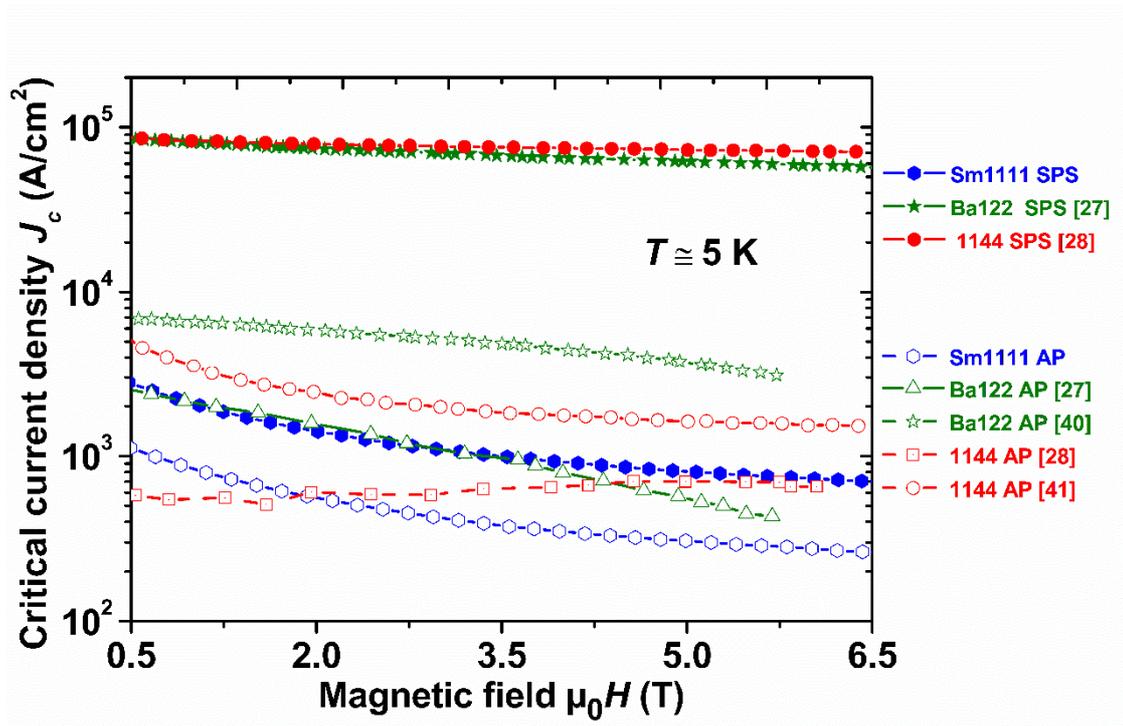

# Effect of spark plasma sintering on the superconducting properties of Sm-based oxypnictide

Mohammad Azam[1], Tatiana Zajarniuk[2], Konrad Kwatek[3], Paolo Mele[4], Shiv J. Singh[1†]

[1]*Institute of High Pressure Physics (IHPP), Polish Academy of Sciences, Sokołowska 29/37, 01-142 Warsaw, Poland*

[2]*Institute of Physics, Polish Academy of Sciences, Aleja Lotników 32/46, 02-668 Warsaw, Poland*

[3]*Faculty of Physics, Warsaw University of Technology, Koszykowa 75, 00-662 Warsaw, Poland*

[4]*College of Engineering, Shibaura Institute of Technology, 307 Fukasaku, Minuma-ku, Saitama 337-8570, Japan*

[†]Corresponding author: Shiv J. Singh

Email: sjs@unipress.waw.pl

https://orcid.org/0000-0001-5769-1787

**Figure S1:** **(i)** Elemental mapping for all constituent elements of SmFeAsO$_{0.80}$F$_{0.20}$ polycrystalline samples prepared by SPS methos: **(i)** SPS-1, **(ii)** SPS-2, **(iii)** SPS-3, **(iv)** SPS-5, **(v)** SPS-6, **(vi)** SPS-8. First and last images of each sample's data set are SEM image and a combined image of all the constituent element, respectively. The rest images depict the elemental mapping of an individual element Sm, Fe, As, O and F. The circles are used to depict the secondary phases, such as SmAs and SmOF (Sm$_2$O$_3$) or inhomogeneity of the constituent elements.

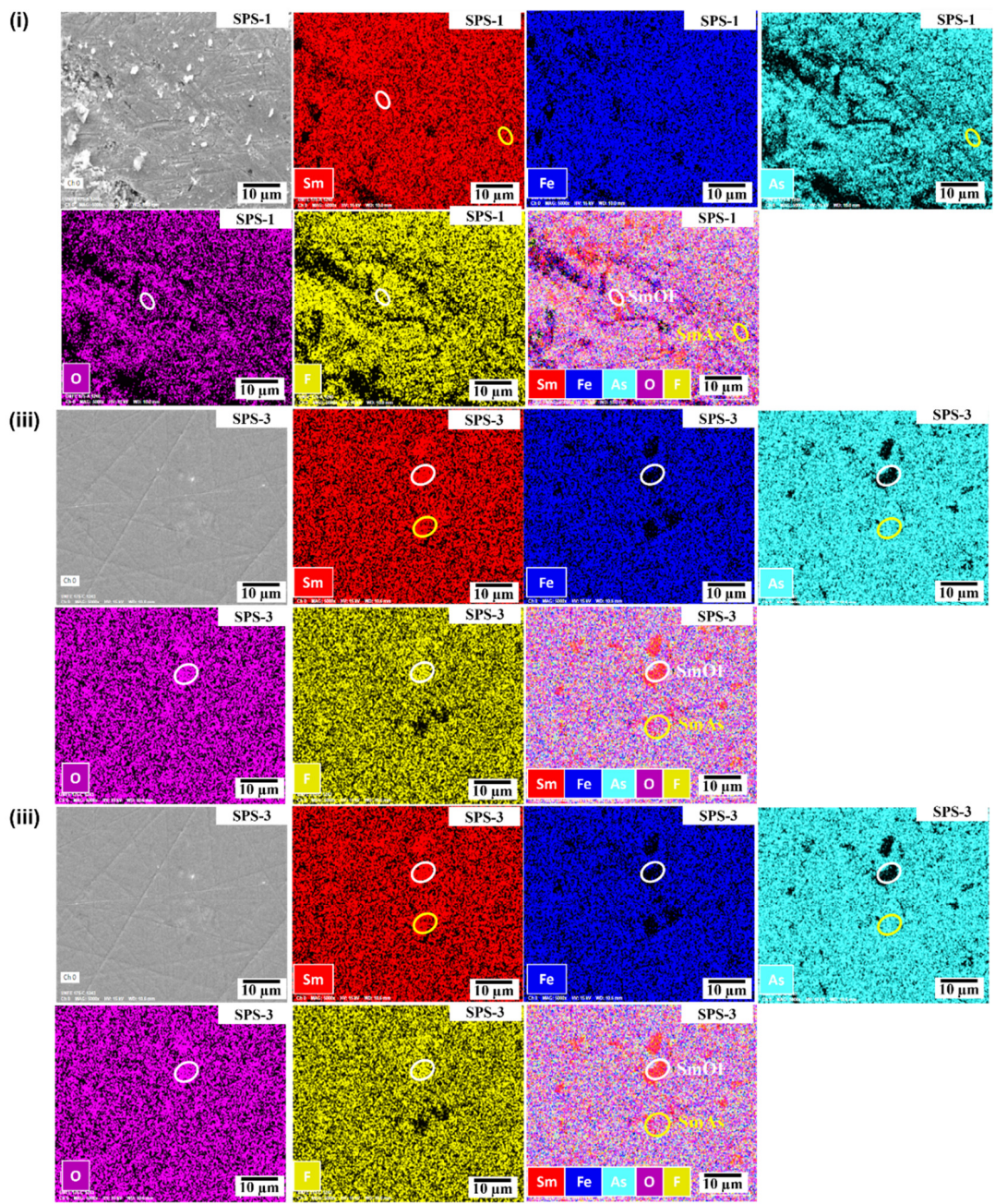

**Figure S2:** Backscattered electron image (BSE; AsB) of **(a-c)** SPS-1, **(d-f)** SPS-2, **(g-i)** SPS-3, **(j-l)** SPS-5, **(m-o)** SPS-6, **(p-r)** SPS-8. Bright, light gray, and black contrast correspond to the phases of $Sm_2O_3$(SmOF) and $SmFeAsO_{0.80}F_{0.20}$ phase, and pores, respectively. The black contrast can be occasionally SmAs.

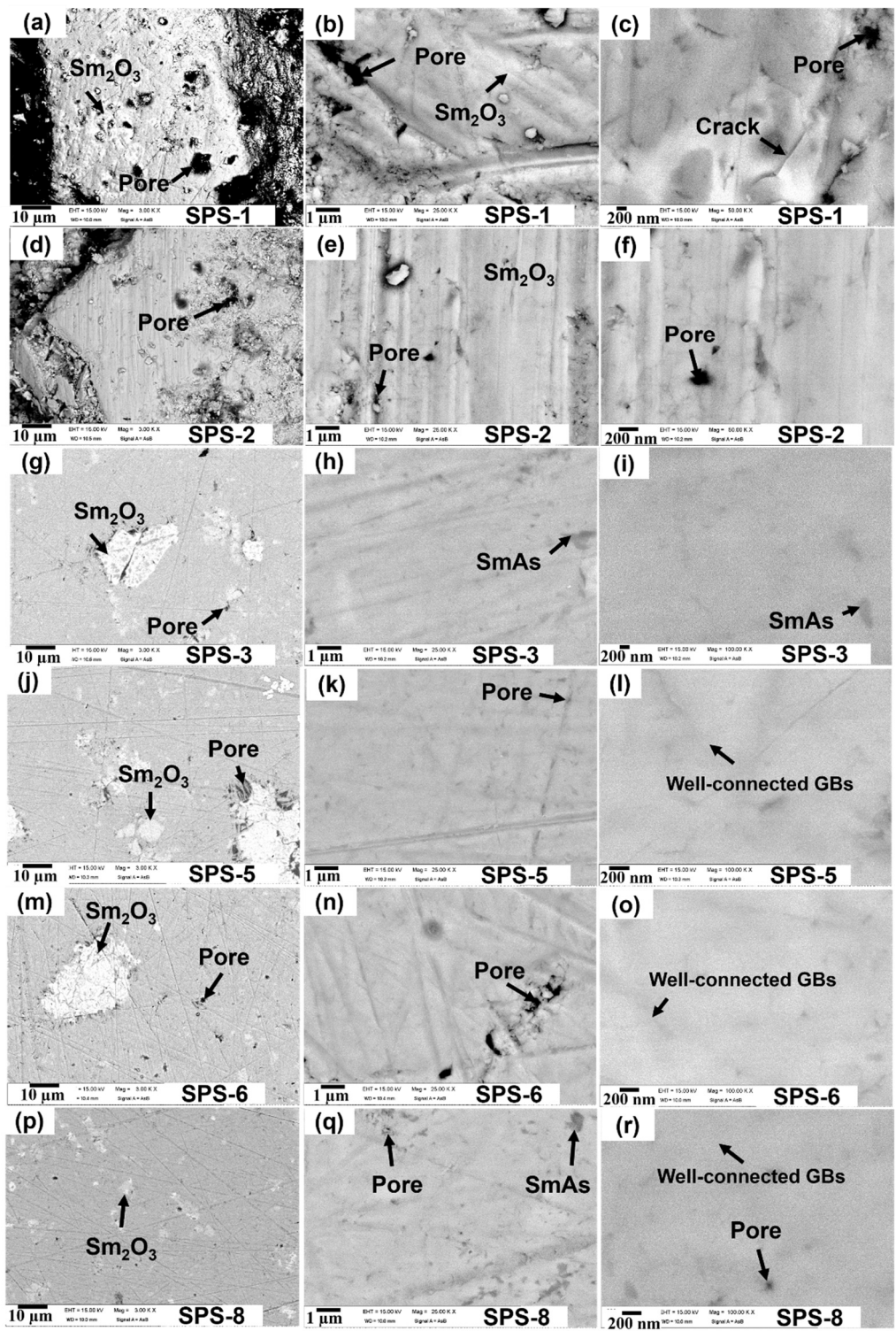

**Figure S3:** The variation of pinning force ($F_p$) at 5 K with an applied magnetic field of up to 8 T for the parent P, HIP, SPS-4, and SPS-7 bulks.

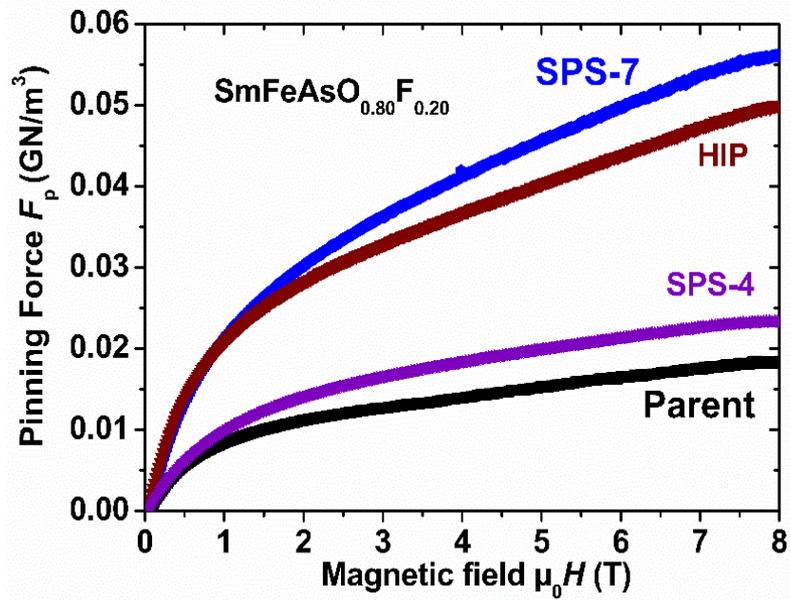

**Table 4:** The onset transition temperature ($T_c$), critical current density ($J_c$), sample density and the impurity phase for F-doped Sm1111 (SmFeAs(O,F): Sm1111), CaKFe$_4$As$_4$ (1144: Ca1144) and (Ba,K)Fe$_2$As$_2$ (Ba122).

| Sample | Properties | Ambient pressure (CSP) | 500 MPa/193 MPa (HP-HTS or HIP) | 45-50 MPa (SPS) |
|---|---|---|---|---|
| **Sm1111** | $T_c$ (K) | ~53 | ~52 | ~51 |
| | $J_c$ (A-cm$^{-2}$) at 0.2 T | $1 \times 10^3$ | $3 \times 10^3$ | $3 \times 10^3$ |
| | Sample density (%) | 50 | 58 | 98 |
| | Impurity phase | Yes | Yes/Same | Yes/Same |
| **Ca1144** | $T_c$ (K) | 33.8 [41] | 35.2 [41] | 35.3 [28] |
| | $J_c$ (A-cm$^{-2}$) at 0.2 T | $5 \times 10^3$ [41] | $\sim 10^4$ [41] | $\sim 10^4$ [28] |
| | Sample density (%) | 66 [41] | 77 [41] | 96 [28] |
| | Impurity phase | Yes | Yes/decreased | Yes/decreased |
| **Ba122** | $T_c$ (K) | 37.7 [40] | 37.1 [40] | 37.8 [27] |
| | $J_c$ (A-cm$^{-2}$) at around 0 - 0.5 T | $\sim 7 \times 10^4$ [40] | $\sim 8 \times 10^4$ [40] | $\sim 10^5$ [27] |
| | Sample density (%) | 68 [40] | 92 [40] | 90 [27] |
| | Impurity phase | Yes | Yes/Decrease | Yes/Decrease |